# Evidence for a Universe Expanding at the Speed of Light

Alasdair Macleod[§]

ABSTRACT
Although big bang cosmology effectively models even the most puzzling observational data, it offers no insight into why the cosmological expansion should occur at all. In this paper it is suggested that a finite Universe poses particular problems at the boundary point when time begins. An alternative model is proposed where the expansion arises from a need to incorporate the boundary effects into observation in a consistent way that avoids discontinuities and singularities. The theory predicts that the Universe is expanding at a constant rate, the speed of light, and is a reasonable match for apparent magnitude – redshift data for both supernovae and 3CR radio sources using only one adjustable parameter, the absolute visual magnitude. Values of –21.8 for the 3CR galaxies and –25.2 for the 3CR quasars give the best fit. The new model is mapped to standard cosmology and predicts a deceleration parameter with a time dependence of the form $q(t) = 1-2t/T$, where $T$ is the current age of the Universe. The picture of expanding space-time with expansion retarded by gravity is usually associated with Friedmann's equation(s) but may be erroneous, giving rise to spurious concepts such as dark energy – in this new model mass and gravity have no effect on the expansion.

*V2.0 Changes: Figure1 changed and minor edits made.*

## 1. Big Bang Cosmology

General relativity is a very elegant and successful theory whose predictions have been verified with great precision for structures of a size comparable with the solar system. It is a natural progression to extend the model and apply Einstein's equations to the entire Universe to predict scale length evolution from mass density. The Friedmann equation is derived by this procedure and describes the dynamic properties of the Universe[1]. The equation has a physical meaning based on the notion of a substantial space-time in which mass is embedded. As space expands, mass is carried with it. The moving mass has kinetic energy that is progressively lost as the expansion works against the gravitational potential. The retardation effect is expected to slow down the expansion over time.

This simple 'big bang' model has actually been shown to be incorrect (or incomplete) from observational data accumulated over the last seven years. It has been discovered that the expansion began to accelerate sometime between 2 and 10 billion years ago and the Universe is still in an acceleration phase today. The Friedmann equation must therefore be modified in an *ad hoc* way to give a scale factor function that possesses the observed properties; this is done by adding a cosmological term representing an unknown large-scale repulsive force or substance referred to as dark energy. The equation also has parameter values that suggest a substantial quantity of dark matter is present in the Universe. The final equation and parameter settings very successfully explain the dynamics of the Universe and all observations. The current status of the 'new cosmology', as it has been dubbed by Turner[2], is satisfactory and the next stage of development is expected to be the identification of dark energy and dark matter and to put constraints on their properties.

In this paper we will look at the philosophical foundations of the Friedmann Equations and the big bang paradigm, and highlight issues that even today, 80 years on, are worthy of exploration. One major objective of a fundamental theory (loosely stated) is to offer a simpler causative explanation for the behaviour under investigation. It should ideally make possible an intuitive understanding of the underlying process. The big bang model does not achieve this, although this is by no means evidence the theory is incorrect – it is however suggestive of a problem. It is certainly the case that the Universe is expanding and had a beginning in time: we are looking to the big bang paradigm to explain why it is expanding at all and why it is expanding at a particular rate at a particular time.

We can ask very specific questions. Why is the rate of expansion of the order of the speed of light? This is a peculiar coincidence. How did the expansion begin? If it is space that is expanding, why does the gravitational effect on the embedded mass reduce the expansion rate rather than lead to proper motion of mass with respect to the expanding space-time manifold? Space and matter must be connected in a rather complex fashion resulting in a metaphysical model of the Universe that is much more complex than before because the actual relationship between matter and space is quite unclear. We expect reciprocal behaviour – if matter movement can affect space expansion, we would expect normal motion through space as a result of conventional forces to give rise to expansion or contraction. This appears not to happen; the necessary coupling between space–time and matter is quite mysterious and certainly not contained in the tenets of general relativity. Finally, why does the expansion not affect bound systems? The big bang model offers no answer to these questions and one is left only with the unsophisticated analogy of an explosion at $T = 0$ (which, to be fair, many cosmologists reject). While impressive in practice, the big bang model is philosophically unsatisfactory.

---

[§] University of the Highlands and Islands, Lews Castle College, Stornoway, Isle of Lewis, Scotland, UK (Alasdair.Macleod@lews.uhi.ac.uk)

Alternative metaphysical models that actually explain the expansion should be welcomed and must be considered in spite of the wide-held view that cosmology is 'solved'[3]. One 'sample' concept will be introduced in this paper and we will take it through the stages of evaluating its philosophical worth to predictions about the behaviour of the Universe and finally comparing these predictions with observation. It is however only one of a class of models, all of which reject the conventional notion that the dynamics of the Universe is described by Friedmann's equation.

**2. A Consistent Metaphysical Model**

To understand the Universe, we can proceed in the manner of Descartes; determine what it is we can be certain of, and construct a model on that foundation. The Universe certainly exists, and though some may disagree, the Universe is consistent[¶]. One facet of this consistency is the conservation laws, primarily energy and momentum. In the context of consistency, the questions about the origin of the cosmological expansion posed in the previous section can be rephrased: *Why is it necessary for the Universe to expand in order that consistency be maintained?* Or equivalently, what inconsistency would arise if the expansion did not occur?

We can try to answer this question from the viewpoint of the observer in the context of what is directly observable. From this perspective, the retarded frame is appropriate[4]. In the retarded frame, the observer is in the privileged position of being at the centre of a finite Universe and its oldest inhabitant. Light that is received has been emitted at earlier times from objects that lie on concentric spherical surfaces tracing the regression of time to an outer surface where $T=0$, the point when the Universe came into existence and time began. The outer surface is the transition from no time (or Universe) to existence and time progression. A major discontinuity can be expected at this point. It is the case in our Universe that the observer always appears protected from singularities and discontinuities. We might therefore expect the Universe will operate in some way to prevent the $T=0$ discontinuity becoming apparent at an observational or operational level. One way this can happen is if the outer surface recedes from the observer at the speed of light. The way this works can be illustrated with an example: Consider the observer at time $T=T_o$. Let the observer be in causal contact with an emitter at $T=0$ at the edge of the Universe, a distance $cT_o$ away. After a time increment $\Delta T$ in the observer's frame, time has moved on to $T_o+\Delta T$. The Universe has also expanded by $c\Delta T$ hence photons received at this later time from the same source will also appear to have originated from time $T=0$. The observer can remain in causal contact with entities for whom time does not progress, thus the transition point as time began to 'flow' is incorporated into the observable world[§]. It is suggested this may be the reason for the cosmological expansion – in the example, if there were no recession velocity then the observer would perceive photons received at $T_o+\Delta T$ to have originated from the source on the boundary at a later time, which of course is impossible if time is not progressing at the boundary - there is no later time. The cosmological expansion is a direct consequence of the finite age of the Universe: mass density and gravitation are irrelevant.

The model is simple and philosophically very appealing because it deals nicely with the beginning of time and incorporates it into observation. A prediction of the new model is that the Universe is expanding at a constant rate $c$ and we may presume intermediate locations scale proportionally (although there is no compelling reason why this should be the case – we must be careful not to be over-influenced by the 'expanding balloon' model of general-relativistic cosmology) . The apparent velocity can be considered equivalent to a proper velocity with radiation subject to the appropriate Doppler shift. The redshift-distance function of this model is thus very different to that of standard cosmology and is readily tested using existing observational data. Although the model is very similar to that of general relativity, there are important differences –for example, the expanding universe is the $T=0$ surface, not us, the observers and space is not expanding; in fact it does not have to exist as this is essentially a relationist model.

---

[¶] In other words, the observer is always presented with a consistent view of the world.

[§] The example is purely illustrative and it is not actually suggested that particles which have no time progression can actually emit photons.

**3. Observational Data**

We consider the simplest form of the apparent velocity equation. This is just an extension of the linear Hubble law over the entire domain, but the distance is not the proper distance but the retarded distance, i.e. the photon journey time or look back time multiplied by *c*. The Hubble constant *H* is simply the reciprocal of the current time. A distant object will have an apparent velocity relative to the observer of [§]

$$v = c \frac{T_o - T_e}{T_o}, \qquad (1)$$

where $T_o$ is the current time and $T_e$ is the emission time[¶]. Throughout the paper, the current age of the Universe is taken as 13.7 Gyr. The current size of the Universe is $cT_o$ and with 71 km/s/Mpc as the present value of the Hubble constant, the boundary is receding at exactly the speed of light, consistent with the assumptions of the model. Space-time is unconditionally flat and the recession velocity is treated special relativistically. The redshift *z* is derived from the Doppler equation for an outward radial proper velocity *v*:

$$z = \sqrt{\frac{1 + v/c}{1 - v/c}} - 1. \qquad (2)$$

The model is readily tested using so-called 'standard candles'. A large set of redshift – apparent magnitude measurements have been accumulated for a variety of object types around $z = 1$. The procedure is to take the measured redshift and extract the equivalent velocity using equation (2). The apparent distance (in parsecs) is then found by taking the difference between the apparent and absolute magnitudes, $m - M$, and using the magnitude-distance formula:

$$5 Log(d) = m - M + 5. \qquad (3)$$

To get the photon travel distance $d_t$, the apparent distance is divided by (1+*z*) (to correct for photon energy loss and the number reduction),

$$d_t = d(1 + z). \qquad (4)$$

The points (*v/c*, $d_t$) are then plotted. From equation (1) the points should fall a straight line that cuts through (0,0) and has a gradient of $cT_o$. There are two classes of cosmological objects that are considered standard candles and are suitable for this test. Type Ia supernovae were shown in an earlier paper[5] to be similar to Einstein-de Sitter cosmology (1). The paper also gives details about the derivation of equations (1) to (4).

In this paper we will consider the radio galaxy 'standard candle'. All the sources except the single BLac from the 1985 update of the Revised Third Cambridge Catalogue (3CR)[6] were processed as described above. The catalogue entries are categorised into quasars, galaxies and N galaxies. The model has only one adjustable parameter for fitting, the absolute magnitude. An absolute visual magnitude of –21.8 was applied for the radio galaxies and –25.2 for the quasars. The points are plotted in *Figure 1* on which the line described by equation (1) is superimposed.

The first point to make is that the radio sources, even the quasars, appear to make reasonable standard candles. The data fit with this new cosmological model is as good as complex attempts with several free variables using standard cosmology[7], and is evidence to support the validity of equation (1) and the new model. However, the quality of the 3CR data is not good enough to make anything other than general predictions so the case for the new model is far from proven.

---

[§] Linearity is just an assumption which should be justified.

[¶] The implication might seem to be that there exists an absolute time which the emitter and absorber reference to 'get' the correct velocity. In fact all the equation is expressing the retarded distance between the emitter and absorber.

**4. Conflict with Standard Cosmology**

Big bang cosmology presumes the size (as represented by scale factor $a(t)$) of the Universe develops according to general relativity with the redshift being a consequence of space expansion[1]. Specifically

$$(1+z) = \frac{a(T_o)}{a(T_e)}. \tag{5}$$

The information about the size of the Universe extracted from the redshift in standard cosmology is very different. The Universe scale factor is rather an unsatisfactory concept to introduce because it is so difficult to measure and is not directly observable – the equations only give us information about the ratio. However, if we take the new model to be a roughly equivalent description of Universe by virtue of the reasonable agreement with data, then using equations (1) and (2) the redshift can be found as a direct function of the age of the Universe. Slotting this information into equation (5), we may get an approximate idea of Universe scales in general-relativistic cosmology. Basically, we are deriving one form of the $a(t)$ function can take to match observation. Using equation (2) for $(1+z)$ and with $t$ in place of $T_e$, equation (1) becomes

$$\frac{a(T_o)}{a(t)} = \sqrt{\frac{1+v/c}{1-v/c}}. \tag{6}$$

$v$ is actually specified by equation (1) and with this substitution the explicit expression for $a(t)$ in standard cosmology will be

$$a(t) = a(T_o)\left(\frac{2T_o}{t} - 1\right)^{\frac{1}{2}}. \tag{7}$$

Differentiating twice;

$$\dot{a}(t) = a(T_o)\frac{T_o}{t^2}\left(\frac{2T_o}{t} - 1\right)^{-\frac{3}{2}}, \tag{8}$$

and

$$\ddot{a}(t) = a(T_o)\frac{T_o}{t^3}\left(2 - \frac{T_o}{t}\right)\left(\frac{2T_o}{t} - 1\right)^{-\frac{5}{2}}. \tag{9}$$

Standard cosmology defines a deceleration parameter $q$ in the following way:

$$q = -\frac{a(t)\ddot{a}(t)}{\dot{a}(t)^2}. \tag{10}$$

Slotting equations (7), (8) and (9) into equation (10),

$$q = 1 - \frac{2t}{T_o}. \tag{11}$$

In the standard model, the expansion will seem to decelerate until time $t = T_o/2$ (about 6.85 billion years ago) when it begins to mysteriously turn and accelerate. Of course, no such thing is really happening –it is an artefact of the equations (equation (5) in particular). The Universe is actually expanding at a steady rate $c$ independent of mass density or gravitation. $a(t)$ is not a meaningful parameter and it does not represent the size of the Universe in the special relativistic model

The situation actually messier – the basic Friedmann equations will never admit a solution for $a(t)$ of the correct form to match cosmological data (i.e. equation (7)), therefore additional terms must be added to force the desired solution. Because the original terms are attributed a physical meaning (kinetic energy, gravitational energy etc), the added terms are also given physical meaning. This then is the origin of dark matter and dark energy; merely artefacts of an incorrect philosophical description of the expansion of the Universe.

Of course, there is no real justification for the adoption of a linear mapping in equation (1). The behaviour at the limits has been constrained by observational data and the nature of the model, but the detailed shape of the function is unknown. It should certainly be special relativistically consistent. The adoption of the linear form is really the powerful 'expanding balloon' concept of general relativity invading the special relativistic model. In the new model space is not expanding – time is growing and the nature of inter-particle interaction is changing with it. The new model does not match supernova data as well as the standard ΛCDM model[5] but equation (1) may be altered to improve the fit and give a much better prediction of $a(t)$ and thus $q$.

**5. Conclusion**

The cosmological expansion thus arises in this new model because time and the Universe had a beginning. Expansion enables the discontinuity at the start point to be correctly managed. The explanation is philosophically superior to the rather crude analogy to an explosion adopted by the big bang model. The new model is in reasonable agreement with observation giving a fit for both supernova and radio galaxy data that suggests the special relativistic model is worthy of further investigation. The idiosyncrasies of standard cosmology (accelerated expansion and dark energy) are correctly identified and roughly quantified. It is implicit that the Universe is a causally connected whole and that there are no disconnected parts. Inflation is therefore discounted and in fact unnecessary as space is unconditionally flat (although space-time will be curved because of the local gravitational force).

The new model does however give rise to a range of new questions. One of these involves energy conservation, which is the cornerstone of the philosophy of the new model. Although gravitation has no effect on the expansion, energy must still be conserved as space expands. It has been shown in another paper in this series[8] that, although a constant expansion rate conserves momentum, gravitational energy is not conserved. One may speculate that the gravitational constant <u>increases</u> with the cosmological expansion. This would certainly result in energy conservation but there is strong evidence the gravitational constant does not vary. However the available data is with respect to the solar system which is not subject to the cosmological expansion; we would not expect a variation in the gravitational constant in this domain.

Another important issue is time. Equation (1) would appear to suggest that there exists a universal absolute time but that is not the case. For each observer, all measurements are relative to their timeline not a universal time. The new model does not deny general relativity, it simply proposes that the true source of the cosmological expansion is the absolute consistency of interaction established by special relativity.

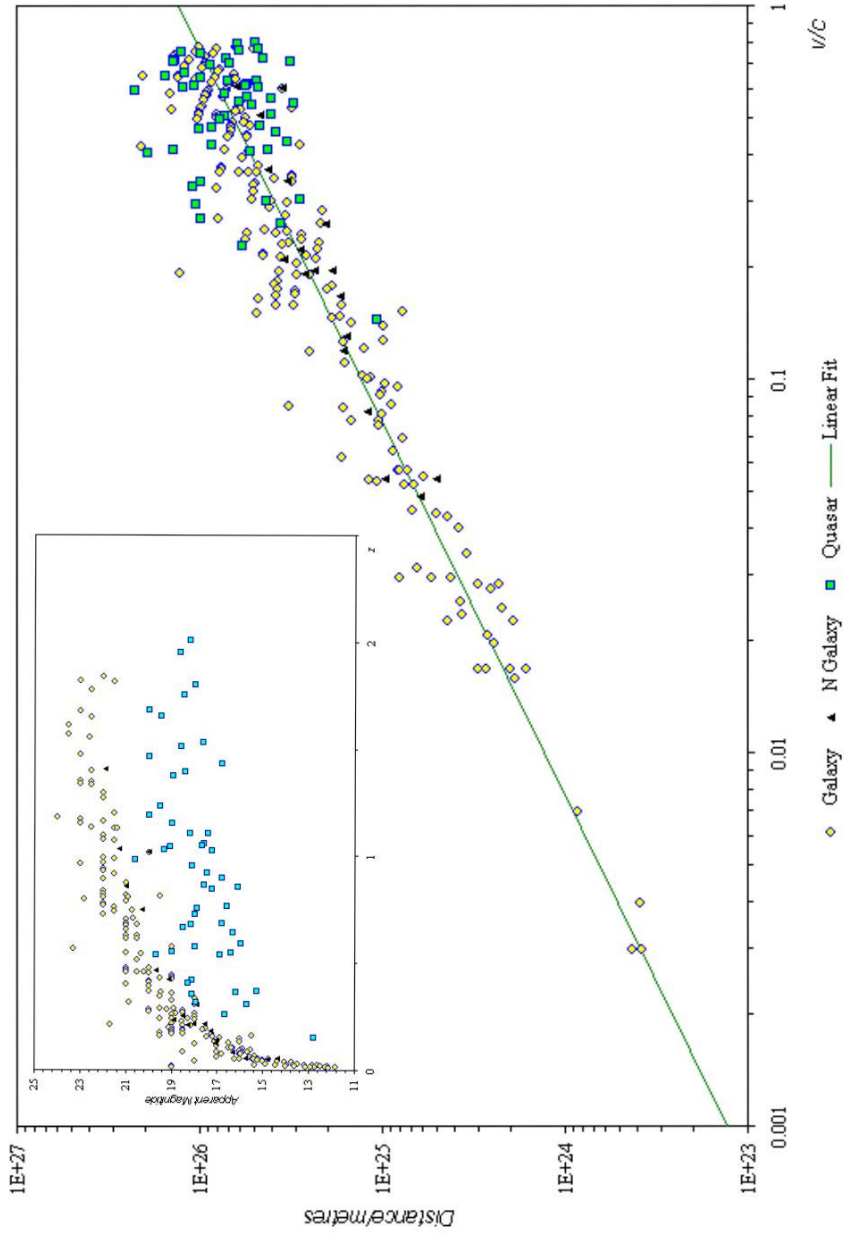

**Figure 1** Doppler velocity- photon travel distance plot for a collection of strong radio sources from z = 0.003 to z = 2. The galaxies were fitted to an absolute visual brightness of -21.8 and the quasars were fitted to -25.2. The solid line is the prediction of equation (1). The inset shows the raw data. For clarity, error bars are not shown.